\documentclass[fleqn,12pt,a4paper]{article}
\usepackage{fullpage}
\usepackage{setspace}
\usepackage{natbib}
\usepackage[hidelinks]{hyperref}
\usepackage{amsfonts}
\usepackage{amssymb}
\usepackage{latexsym, amsmath, amsthm, amssymb, natbib, verbatim,subfigure}
\usepackage[monochrome]{color}
\usepackage{graphicx}
\usepackage{soul}
\usepackage{caption}
\usepackage{floatpag}
\usepackage[colorinlistoftodos]{todonotes}

\newtheorem{proposition}{Proposition}

\newtheorem{definition}{Definition}

\newtheorem{remark}{Remark}

\newtheorem*{claim*}{Claim}
\newtheorem*{corollary*}{Corollary}

\theoremstyle{remark}
\newcommand{\reals}{\mathbb{R}}

\title{\vspace{-40pt}\textbf{Non-Diversified Portfolios with Subjective 
		Expected Utility}
	}

\author{
	\ \\ Christopher P. Chambers\thanks{\href{mailto:Christopher.Chambers@georgetown.edu}{Christopher.Chambers@georgetown.edu}} 
	\\ {\footnotesize Georgetown University}  \and 
	\ \\  Georgios Gerasimou\thanks{\href{mailto:Georgios.Gerasimou@glasgow.ac.uk}{Georgios.Gerasimou@glasgow.ac.uk}} 
	\\ {\footnotesize University of Glasgow} \vspace{10pt}}

\date{\ \\ \normalsize \today \vspace{-10pt}}

\begin{document}

\maketitle

\setcounter{page}{0}

\thispagestyle{empty}

\begin{abstract}
Diversification is the typical investment strategy of risk-averse agents.
However, non-diversified positions that allocate all resources to a single asset, 
state of the world or revenue stream are common too. 
We show that whenever finitely many non-diversified demands under uncertainty are 
compatible with risk-averse subjective expected utility maximization 
under strictly positive beliefs, they are also rationalizable under 
the same beliefs by many qualitatively distinct risk-averse as well as 
risk-neutral and risk-seeking preferences.\\

\noindent Keywords: Investment under uncertainty; Non-diversification; 
Subjective expected utility; Demand; Revealed preference.
\end{abstract}

\vfill

\pagebreak

\begin{quotation}
	
	\footnotesize
	
	\hfill \textit{``But the wise man saith,}
	
	\hfill \textit{`Put all your eggs in the one basket and}  
	
	\hfill \textit{ - WATCH THAT BASKET'.''}
	
	\hfill Mark Twain\footnote{Source: \textit{Pudd'nhead Wilson}, 
		Charles L. Webster \& Co, 1894.}
	
	\large
\end{quotation}

\begin{quotation}

\footnotesize

  \hfill \textit{``Diversification is protection against ignorance,}
  
  \hfill \textit{but if you don’t feel ignorant,}
  
  \hfill \textit{the need for it goes down drastically.''}

  \hfill Warren Buffett\footnote{Source: \textit{Warren Buffett: 
  		The \$59 Billion Philanthropist}, Forbes Media, 2018.}

  \large
\end{quotation}

\section{Introduction}

Risk-averse decision makers in real-world and experimental 
markets typically allocate 
their available funds or revenue streams 
in ways that exhibit diversification. 
Yet at the same time it is not uncommon 
for individuals operating in such environments 
to choose non-diversified portfolios that allocate 
all resources to a single asset or 
state of the world.\footnote{From the 207 
experimental subjects in \cite*{halevy-persitz-zrill18}, for example, 
45\% made such a non-diversified demand over Arrow-Debreu 
securities at least once, 11\% did so in at least half of 
their 22 decisions, while the overall rate of such behavior 
was 16\%. For the 93 subjects in \cite*{choi-fisman-gale-kariv07} 
these figures were similar at 51\%, 8.6\% and 11\%, respectively, even though 
these subjects made 50 decisions instead 
(more details are available in our online supplementary appendix).
For a theoretical account on 
how investing in un-diversified portfolios 
could be observed in the presence of financial complexity
see \cite{galanis18}.}

Considering the intuitive link between risk aversion 
and diversification,\footnote{For example, because firm CEOs are often 
considered averse to non-diversified revenue streams, some firm boards 
provide more \textit{``risk-taking incentives''} in the CEOs' compensation 
packages \textit{``to offset their risk of non-diversified revenue streams,
thereby preventing excessive managerial conservatism at the expense
of value maximization''} \citep*{ceo22}.} 
the first question that arises naturally is whether agents who 
have been observed to make such \textit{non}-diversified choices 
can also be portrayed as risk-\textit{averse} 
subjective expected utility (SEU; Savage, \citeyear{savage54}) 
maximizers under some beliefs and tastes. 
The answer to this question is positive and implicitly follows from the 
analysis of \cite{inada63}. In particular, such behaviour is compatible 
with a risk-averse SEU agent, but only if her marginal utility 
is bounded above at all non-negative wealth levels; 
hence, if the respective \cite{inada63} condition on marginal 
utility is violated.

In light of this fact then, we ask:
What is it possible for the analyst to infer from 
a non-diversifying SEU agent's demand under uncertainty 
about their possible attitudes toward risk?
We answer this question by showing that if a finite dataset consisting of 
non-diversified positions is compatible with strictly risk-averse 
SEU under some full-support beliefs, then there is in fact a very general 
class of preferences over wealth that feature bounded marginal 
utility and also rationalize this dataset under the 
\textit{same} beliefs. In particular, we show 
that such a simultaneous rationalization is achievable 
with risk-neutral, risk-seeking, 
constant, as well as increasing absolute risk aversion 
--although not constant relative risk aversion--
preferences. 
Thus, when non-diversified choice behavior is SEU-rationalizable by some 
strictly concave utility function and full-support beliefs, 
there is a precise sense 
in which this behavior is largely uninformative about the decision maker's 
risk attitudes conditional on those beliefs. 

\section{Analysis}

$\Omega:=\{\omega_1,\ldots,\omega_n\}$ is a finite set of states, 
with generic element
$\omega\in \Omega$, and $\pi$ is a probability measure over $\Omega$. 
$\mathcal{D}=\{(p^i,x^i)\}_{i=1}^k$ is a finite dataset of prices 
and asset demands, 
where $p^i,x^i\in\mathbb{R}^n_{+}$ and $p^i\gg \mathbf{0}$ 
for all $i\leq k$.\footnote{
	$x \gg y$ means that $x_{\omega}>y_{\omega}$ for all $\omega \in\Omega$.} 
We will refer to any $x^i$ in $\mathcal{D}$ as a \textit{non-diversified demand} 
if there exists $\omega\in\Omega$ for which $x^i_{\omega}>0$ 
and $x^i_{\omega'}=0$ for all $\omega'\neq \omega$.

\begin{definition}
A dataset $\mathcal{D}=\{(p^i,x^i)\}_{i=1}^k$ is rationalizable by subjective 
expected utility \textnormal{(SEU-rationalizable)} if there is a probability 
measure $\pi$ over $\Omega$ and an increasing function 
$u:\mathbb{R}\rightarrow\mathbb{R}$ such that, for all $i\leq k$,
\begin{equation*}
	\mathbb{E}_\pi u(x^i_\omega) \geq \mathbb{E}_\pi u(x_\omega) 
	\text{ for all } x\in \mathbb{R}^n_{+} 
	\text{ that satisfies } p^i\cdot x \leq p^i \cdot x^i
\end{equation*}
\end{definition}

\begin{remark} \label{rem-1}
	If an SEU-rationalizable dataset contains a non-diversified demand, 
	then it is not rationalizable by a Constant Relative Risk Aversion 
	(CRRA) utility function defined by $u(x):=x^\alpha$ for some 
	$\alpha\in (0,1)$.
\end{remark}

\noindent This claim implicitly follows from \cite{inada63}
and generalizes to any class of utility functions with 
unbounded marginal utility at zero wealth.

We now recall that for any function $u:\reals\rightarrow\reals$, a 
\textit{supergradient} at a point $\overline{x}$ is an element 
$y\in\reals$ for which 
$$u(x)\leq u(\overline{x}) + y(x-\overline{x})$$
holds for all $x\in \reals$. If a point has a single supergradient, then 
that supergradient is its derivative.  The \textit{superdifferential} 
at $\overline{x}$ is denoted by $\partial u(\overline{x})$ and consists 
of all supergradients at $\overline{x}$.

The next definition introduces the class of models that we take interest in. 
We envision a model as a class of utility functions which is ``closed'' 
under certain operations.
Importantly, this class need not be globally increasing in wealth: 
our first requirement is only 
that there exists a function in this class which is strictly increasing 
in a neighborhood of zero
(the relevance of this will be shown below). 
Our second requirement is that this 
neighborhood can be made arbitrarily large.

\begin{definition}\label{def:scalable}
A collection $\mathcal{U}$ of concave and continuous functions from 
$\reals_+$ to $\reals$ is \textnormal{scalable} if it has the 
following properties:
\begin{enumerate}
\item There is $u\in\mathcal{U}$ for which there is 
a supergradient $u'(0)$ at $0$, 
so that for all $x\in \reals$, $u(x)\leq u(0) + u'(0)x$; 
further, all supergradients of $u$ at $0$ are strictly positive.
\item For all $\kappa \in (0,1]$ and all $v\in\mathcal{U}$, $v_{\kappa}$ 
defined as $v_{\kappa}(x):=v(\kappa x)$ 
for all $x\in \reals$ satisfies $v_{\kappa}\in\mathcal{U}$.
\end{enumerate}	
\end{definition}

\begin{remark}
	As supergradients are real-valued, the first part of Definition 
	\ref{def:scalable} implies that the relevant $u\in\mathcal{U}$ 
	must satisfy $u'(0)<\infty$, thereby satisfying the
	premise of Remark \ref{rem-1}.
\end{remark}

\begin{proposition}\label{prop:concave}
	Suppose that a dataset $\mathcal{D}=\{(x^i,p^i)\}_{i=1}^k$ 
	is SEU-rationalizable by a strictly concave, strictly 
	increasing utility index and a 
	full-support probability measure $\pi$, 
	and that each $x^i$ is a non-diversified 
	demand. Then, for any scalable family of concave 
	and continuous utility functions 
	$\mathcal{U}$ there is $u\in\mathcal{U}$ such 
	that $u$ is an SEU rationalization 
	of the dataset under $\pi$.
\end{proposition}

\begin{proof}Let $v$ be the utility index rationalizing the data.  
	Without loss, we may assume that $v(0)=0$.  
	Let $x^i$ be such that coordinate $x^i_{\omega^i}>0$, 
	and all remaining coordinates are zero.  
	Slater's condition is satisfied here, 
	so by Theorems 28.2 and 28.3 of \cite{rockafellar70}, 
	this implies that there is a supergradient $y^i_{\omega}$ of $v$ 
	for each $\omega\neq\omega^i$ at $0$ and a supergradient $y^i_{\omega^i}$ 
	at $x^i_{\omega^i}$, and a multiplier $\lambda^i>0$ for which 
	$$\pi(\omega) y^i_{\omega} - \lambda^ip^i(\omega)\leq 
	\pi(\omega^i)y^i_{\omega^i}-\lambda^ip^i(\omega^i)=0.$$ 
	[The fact that supergradients are additive follows from 
	Theorem 23.8 in \cite{rockafellar70}.]  
	Each $y^i_{\omega}>0$ as $v$ is strictly increasing. 
	We may conclude then that for all 
	$\omega \neq \omega^i$, $\lambda^ip^i(\omega) \geq \pi(\omega)y^i_{\omega}$ 
	and that 
	$\lambda^i p^i(\omega^i) = \pi(\omega^i)y^i_{\omega^i}$.  
	So 
	\begin{equation}\label{eq:inequality}
		\frac{ p^i(\omega^i)}{ p^i(\omega)}\leq 
		\frac{y^i_{\omega^i}\pi(\omega^i)}{y^i_{\omega}\pi(\omega)}
		<
		\frac{\pi(\omega^i)}{\pi(\omega)},
	\end{equation} 
	where the strict inequality follows from strict concavity of $v$ 
	and the fact that the superdifferential is strictly decreasing.

Now, fix any $u\in\mathcal{U}$ with finite supergradient at $0$, 
whose supergradients are all strictly positive there. 
Without loss, suppose that $u(0)=0$.  
Let $u'(0)$ denote the minimal such supergradient 
(the one with the smallest value); 
the set of supergradients (the superdifferential) is well-known to be closed 
(see p. 215 in Rockafellar, \citeyear{rockafellar70}), so such an element exists.  
Without loss assume $u'(0)=1$ (this is possible because $\mathcal{U}$ is scalable). 
If the superdifferential correspondence is constant and equal to $u'(0)$, 
no more work is needed (this means that $u$ is a linear function). 
Otherwise, we claim that for any $\epsilon > 0$, there exists $x^*>0$ 
with a supergradient bounded below by $1-\epsilon$.  
To see why, observe that if $x_n \rightarrow 0$ strictly monotonically 
and $y_n \in \partial u(x_n)$, then $y_n$ is weakly increasing and thus 
has a limit; the limit must be a member of $\partial u(0)$ 
by Theorem 24.4 of \cite{rockafellar70}, and hence must be at least as 
large as $1$ (as $1$ was the minimal element of $\partial u(0)$).  
Consequently there is $x_n>0$ small so that $y_n$ is a supergradient 
of $u$ at $x_n$ and $y_n \geq 1-\epsilon$, which is what we wanted to show.  
Obviously, $y_n \leq 1$.

Now choose $\epsilon>0$ small so that, for all $i$, 
we have 
$\frac{(1-\epsilon)\pi(\omega^i)}{\pi(\omega)}
>
\frac{p^i(\omega^i)}{p^i(\omega)}$; 
this can be done by finiteness of the set of observations.  
Let $\overline{x}=\max_i x^i_{\omega^i}$ 
be the maximal nonzero consumed commodity; 
obviously $\overline{x}>0$.  
Let $0<x_{\epsilon}<\overline{x}$ have a supergradient of at least $1-\epsilon$.  
Let $\alpha = \frac{x_{\epsilon}}{\overline{x}}<1$, 
so that for all $i$, $\alpha x^i_{\omega^i} < x_{\epsilon}$.  
Observe that $u_{\alpha}(\overline{x})=u(x_{\epsilon})$.

Now, define $\overline{u}(x) = \frac{u_{\alpha}(x)}{\alpha}$.  
By assumption, $u_{\alpha}\in\mathcal{U}$, and since $\overline{u}$ is 
cardinally equivalent to $u_{\alpha}$, they have the same optimizers in 
any constrained optimization problem.  

Observe that $1\in\partial \overline{u}(0)$ and that there is a supergradient 
of $\overline{u}$ at $x_{\epsilon}$ at least as large as $1-\epsilon$.  
Therefore for each $i$, $\overline{u}$ has a supergradient at $x^i_{\omega^i}$ 
at least as large as $1-\epsilon$, as $\alpha x^i_{\omega^i}<x_{\epsilon}$.  
Consequently, by letting $z^i_{\omega^i}$ be any member of the supergradient 
of $x^i_{\omega^i}$ at least as large as $1-\epsilon$, we have 
$$
\frac{ p^i(\omega^i)}{ p^i(\omega)}
< 
\frac{z^i_{\omega^i}\pi(\omega^i)}{\pi(\omega)}.
$$  
Set $\lambda^i = \frac{\pi(\omega^i)z^i_{\omega^i}}{p^i(\omega^i)}$ 
and observe that we 
then have $\pi(\omega^i)z^i_{\omega^i}-\lambda^i p^i(\omega^i) = 0$ 
and $\pi(\omega)-\lambda^i p^i(\omega)<0$.  
Conclude again by Theorem 28.3 of \cite{rockafellar70}, 
using the fact that $1$ is a supergradient of $\overline{u}$ at $0$.
\end{proof}

Proposition \ref{prop:concave} can be extended to the risk-seeking case by 
adapting Definition \ref{def:scalable}.

\begin{definition}
A class $\mathcal{U}^*$ of continuous, increasing and convex functions 
from $\reals_+$ to $\reals$ is \textnormal{scalable} if it has the 
following properties:
\begin{enumerate}
	\item There is $v\in\mathcal{U}^*$ with a positive subgradient at $0$.
	\item For all $\alpha\in (0,1]$ and all $v\in\mathcal{U}^*$, $v_{\alpha}$ 
	defined as $v_{\alpha}(x):=v(\alpha x)$ 
	for all $x\in \reals$ satisfies $v_{\alpha}\in\mathcal{U}^*$.
\end{enumerate}
\end{definition}

\begin{proposition}\label{prop:convex} 
	Suppose that a dataset $\{(x^i,p^i)\}_{i=1}^k$ is SEU-rationalizable 
	by a strictly concave utility function and a full-support probability 
	measure $\pi$, and that each $x^i$ is a non-diversified demand. 
	Then, for any scalable family of increasing, convex, 
	and continuous utility functions $\mathcal{U}^*$ 
	there is some  $u\in\mathcal{U}^*$ such that $u$ 
	is an SEU rationalization of the dataset under $\pi$.
\end{proposition}

\begin{proof}
	Observe that \eqref{eq:inequality} implies 
	$\frac{\pi(\omega^i)}{p^i(\omega^i)}>\frac{\pi(\omega)}{p^i(\omega)}$ 
	for any $\omega\neq\omega^i$.  Consequently, the linear utility given 
	by $v(x) = \sum_{\omega}\pi(\omega)x(\omega)$ is maximized uniquely 
	at $x^i$ on the budget $\{x:p^i \cdot x \leq p^i \cdot x^i\}$.

	The argument is roughly the same as the preceding one, 
	so we only sketch the remaining.  
	Choose $\alpha>0$ so that (a suitably normalized) $u$ has a 
	subgradient of $1$ at $0$, and so that for a given $\epsilon > 0$, 
	there is a subgradient of $1+\frac{\epsilon}{\overline{x}}$, 
	where again $\overline{x}$ is the maximal observed consumption bundle.  
	Now choose $\epsilon > 0$ so that 
	\begin{eqnarray*}
	\pi(\omega^i)u_{\alpha}(x^i) 
	& = 
	& \pi(\omega^i)u_{\alpha}\left(\frac{1}{p^i(\omega^i)}\right)\\
	&\geq& \frac{\pi(\omega^i)}{p^i(\omega^i)}\\
	& > & \frac{\pi(\omega)(1+\epsilon)}{p^i(\omega^i)}\\
	&\geq& \pi(\omega)u_{\alpha}\left(\frac{1}{p^i(\omega^i)}\right).
	\end{eqnarray*}
	Since a (continuous) convex function is always maximized at an extreme point, 
	by Bauer's Maximum Principle (Theorem 7.69 in Aliprantis and Border, 
	\citeyear{aliprantis-border06}), the result follows. 
\end{proof}

We illustrate the economic relevance of these results 
with the following Corollary, which lists several classes of 
scalable families of utility functions, including the quadratic family, 
(iii.), which is strictly increasing only in a neighborhood of the origin.

\begin{corollary*}\label{corollary}
	If the dataset $\mathcal{D}:=\{(x^i,p^i)\}_{i=1}^k$ is SEU-rationalizable 
	by a strictly concave and strictly increasing utility function under a 
	full-support probability measure $\pi$ and each $x^i$ is non-diversified, 
	then $\mathcal{D}$ is also SEU-rationalizable under $\pi$ by a:
	\begin{enumerate}
        \item[(i)] $u_1^{\alpha}$, for any $\alpha \in (0,1)$, such that for 
        some $c_\alpha>0$, $u_1^{\alpha}(x):=(x+c_\alpha)^\alpha$\\ 
        (DARA\footnote{D(C)(I)ARA refers 
        		to decreasing (constant) (increasing) 
        		absolute risk aversion.}-risk-averse 
        		with positive fixed initial wealth);\footnote{When 
        	$c_\alpha$ is negative, it is also known 
        	as the \textit{subsistence parameter} \citep{ogaki-zhang01}.}
		\item[(ii)] $u_2$ such that, 
		for some $\beta>0$, $u_2(x):=1-e^{-\beta x}$\\ (CARA-risk-averse);
  		\item[(iii)] \label{it:quadratic} $u_3$ such that, 
  		for some $\lambda>0$, 
  		$u_4(x):=x-\lambda x^2$\\ (IARA-risk-averse/increasing 
  		in a neighborhood of 0);
        \item[(iv)] $u_4$ such that, for some $\gamma>0$, 
        	$u_3(x):=\dfrac{x}{1+\gamma x}$\\ 
        	(IARA-risk-averse/increasing in $\mathbb{R}_+$);
		\item[(v)] Linear $u$;
		\item[(vi)] Strictly convex $v$.
	\end{enumerate}
\end{corollary*}

\begin{proof}
	We apply Proposition \ref{prop:concave} separately to the first four cases. 
	It is immediate that the last two satisfy the conditions of 
	Proposition \ref{prop:convex} too, and that many suitable classes of 
	functions can be constructed for the last case.
	
	\noindent (i) Fix $\alpha\in (0,1)$.  
	Let $\mathcal{U}$ denote the set of all utility indices $u$ 
	for which there exists $\kappa > 0$ such that 
	$u(x) = (1+\kappa x)^{\alpha}$. 
	Then it is obvious that $\mathcal{U}$ is scalable. 
	Further, fixing $\kappa = 1$, say, $u(x) = (1+x)^{\alpha}$ 
	is concave and continuous, there is a supergradient at $0$, 
	and all supergradients are strictly positive. 
	By Proposition~\ref{prop:concave}, there is $\kappa > 0$ 
	for which $u(x) = (1+\kappa x)^{\alpha}$ rationalizes the data. 
	The result concludes by observing that this utility index 
	is cardinally equivalent to $v(x) = (\kappa^{-1} + x)^{\alpha}$, 
	where we then set $c_{\alpha}=\kappa^{-1}$.
	
	\noindent (ii) Let $\mathcal{U}$ denote the set of functions $u$ 
	for which there exists $\beta > 0$ so that $u(x) = 1-e^{-\beta x}$ 
	and observe that this family is scalable.
	
	\noindent (iii) Observe that the family $\mathcal{U}$ defined by 
	$u\in\mathcal{U}$ if there exist $\theta > 0$ and $\mu>0$ 
	for which $u(x) = \theta x -\mu x^2$ is scalable.  Finally, 
	each such $u\in\mathcal{U}$ is cardinally equivalent to 
	$v(x) = x -\frac{\mu}{\theta}x^2$, from which the result follows.
	
	\noindent (iv)  Observe that the family $\mathcal{U}$ defined by 
	$u\in \mathcal{U}$ if there exists $\beta,\gamma > 0$ for which 
	$u(x) = \dfrac{\beta x}{1+\gamma x}$ is scalable. Finally, each 
	such $u\in\mathcal{U}$ is cardinally 
	equivalent to $v(x)=\dfrac{x}{1+\gamma x}$.
\end{proof}

\begin{figure}[htpb!]
	 \floatpagestyle{empty}
	\caption{\centering Example dataset with non-diversified 
		demands that affords SEU rationalizations under 
		the same beliefs and 6 distinct attitudes to risk.}
	\centering
	\includegraphics[width=.75\textwidth]{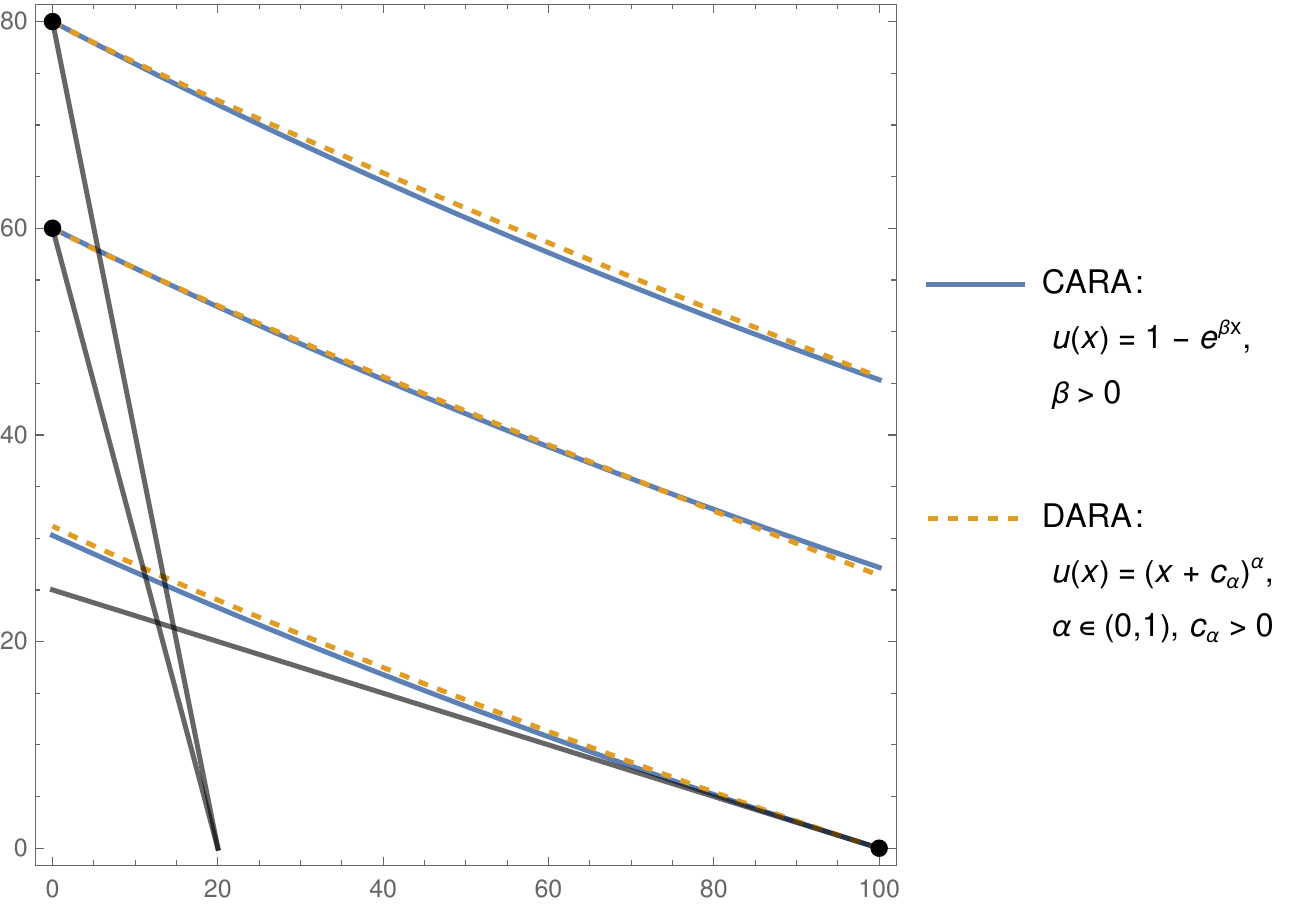}
	\includegraphics[width=.75\textwidth]{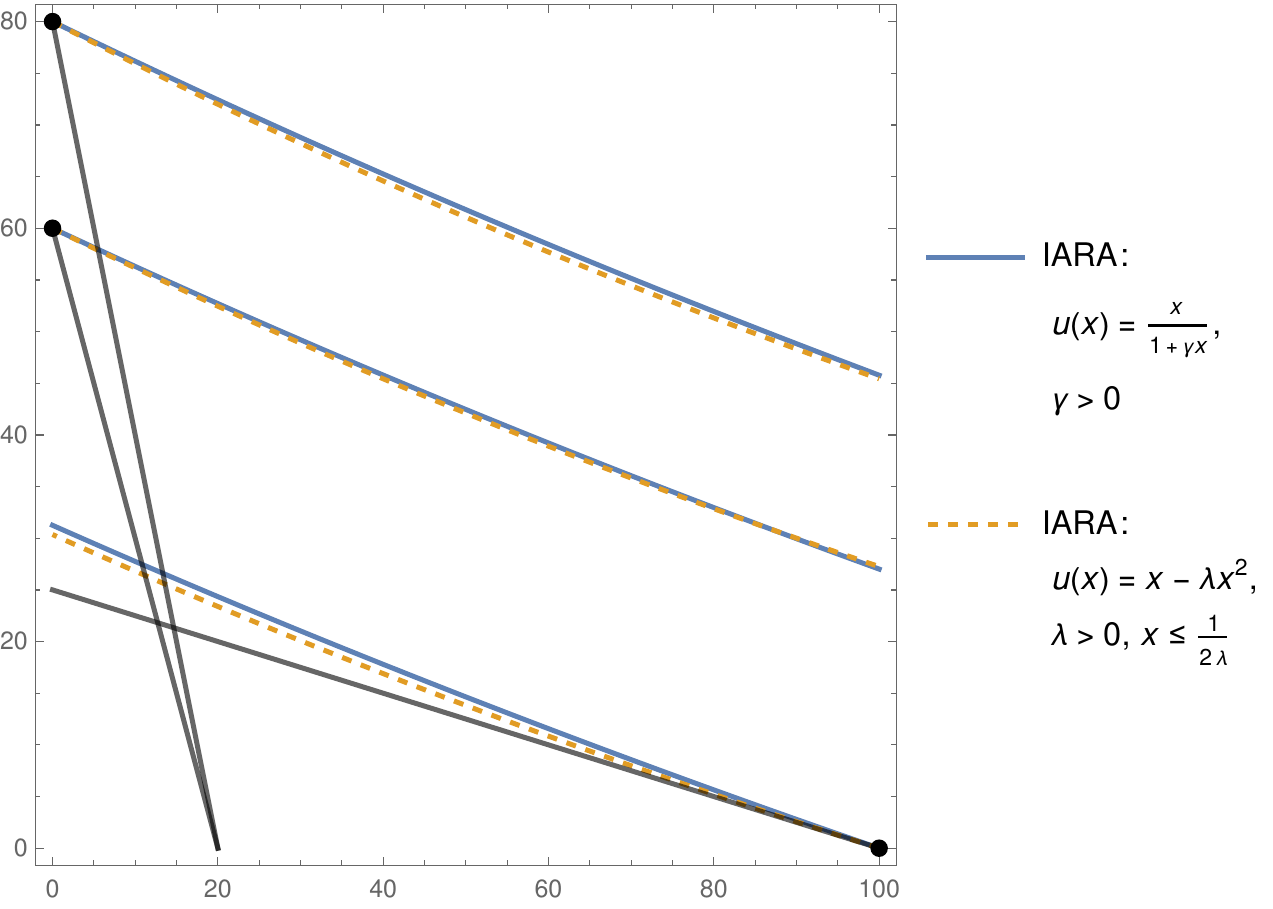}
	\includegraphics[width=.75\textwidth]{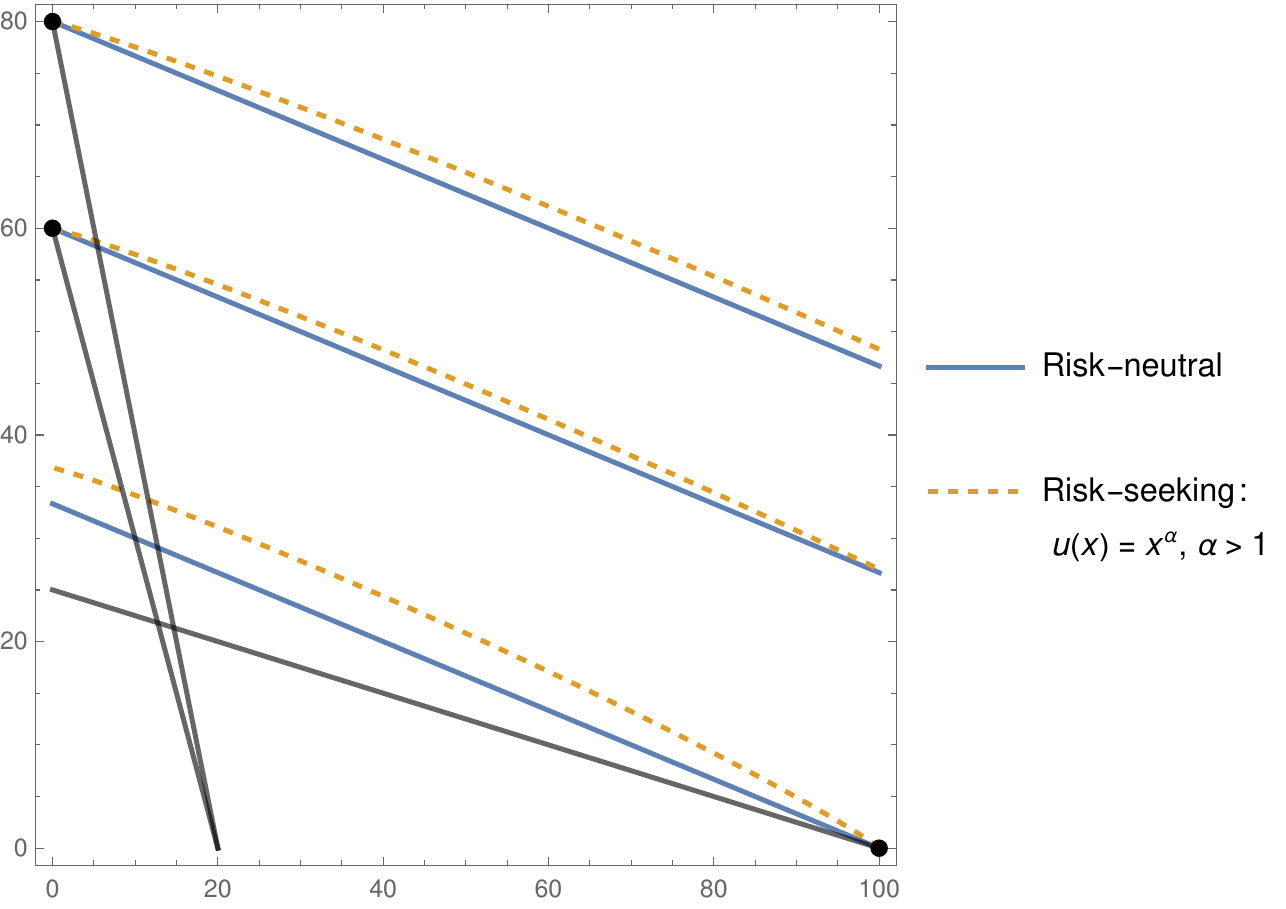}
	\label{fig:example}
\end{figure}

\section{Example}

Assume two states of the world and consider the non-diversified-demand 
example dataset 
$$\mathcal{D}:=\left\{\underbrace{\big((100,0),(1,4)\bigr)}_{(x^1,\, p^1)},
\underbrace{\big((0,80),(4,1)\bigr)}_{(x^2,\, p^2)},
\underbrace{\big((0,60),(3,1)\bigr)}_{(x^3,\, p^3)}\right\}.$$
It is easy to see that $\mathcal{D}$ satisfies the 
Generalized Axiom of Revealed Preference (GARP) 
and is therefore rationalizable by some ordinal utility index \citep{afriat67}. 
Furthermore, for the two relevant sequences $(x^1,x^2)$ and $(x^1,x^3)$ in our 
example we observe that, for $j\in\{2,3\}$, $x^1_1>x^j_1$, 
$x^j_2>x^1_2$, $x^1_1>x^1_2$, $x^j_2>x^j_1$ 
and $\frac{p^1_1p^j_2}{p^1_2p^j_1}<1$. 
Hence, the SARSEU axiom of \cite{echenique-saito15} is satisfied and, 
by Theorem 1 in that paper, 
there exists a concave and strictly increasing utility 
index $u$ and a full-support probability measure 
$\pi$ on $\Omega:=\{1,2\}$ that form a risk-averse 
SEU rationalization of $\mathcal{D}$. 

We now illustrate our main results on this example dataset for the 
fixed beliefs $\pi:=\left(\frac{1}{4},\frac{3}{4}\right)$ 
(see also Fig. \ref{fig:example}). With a utility index $u$, 
optimality of choice $x^i$ at prices $p^i$, $i\leq 3$, is equivalent 
to the marginal rate of substitution condition
\begin{eqnarray}\label{optimality}
	\frac{\pi_{\omega^i}}{\pi_{\omega'}}\frac{u'(x^i_{\omega^i})}{u'(0)} 
	& 
	\geq 
	& \frac{p^i_{\omega^i}}{p^i_{\omega'}}
\end{eqnarray}
In case (i) with $u$ featuring $c_\alpha=1$ these conditions become 
$\frac{1}{3}\cdot 101^{\alpha-1}\geq \frac{1}{4}$, 
$3\cdot 81^{\alpha-1}\geq\frac{1}{4}$ 
and $3\cdot 61^{\alpha-1}\geq\frac{1}{3}$, 
and are simultaneously satisfied for 
$\alpha\in(0.9378,1)$, for example. 
In the CARA case (ii), the conditions reduce to 
$\frac{1}{3}e^{-100\beta}\geq \frac{1}{4}$, $3e^{-80\beta}\geq\frac{1}{4}$ and 
$3e^{-60\beta}\geq\frac{1}{3}$. 
These inequalities are satisfied when $\beta<0.00285$, for example. 
For the quadratic-utility IARA index in (iii), 
formulated as $u(x)=\theta x - \lambda x^2$, 
the conditions are $\frac{1}{3}\frac{\theta-200\lambda}{\theta}
\geq 
\frac{1}{4}$, 
$3\frac{\theta-160\lambda}{\theta}\geq \frac{1}{4}$ 
and 
$3\frac{\theta-120\lambda}{\theta}\geq \frac{1}{3}$. 
These conditions are satisfied for any $\theta,\lambda$ 
such that $\theta\geq 800\lambda$.
For the IARA index in (iv) the conditions become 
$\frac{1}{3(1+100\gamma)^2}\geq \frac{1}{4}$, 
$\frac{3}{(1+80\gamma)^2}\geq \frac{1}{4}$, 
$\frac{3}{(1+60\gamma)^2}\geq \frac{1}{3}$ and are satisfied for all 
$\gamma\in \left(0, \frac{2 \sqrt{3}-3}{300}\right]$. 
In the risk-neutral and risk-seeking cases 
(v) and (vi) with linear and strictly convex utility functions, 
finally, \eqref{optimality} is trivially satisfied because 
$\frac{u'(x^i_{\omega^i})}{u'(0)}\geq 1$ 
and $\frac{\pi_{\omega^i}}{\pi_{\omega'}}
\geq 
\frac{p^i_{\omega^i}}{p^i_{\omega'}}$ for all $i$.\hfill $\lozenge$

\bibliographystyle{aer}
\bibliography{corner}

\end{document}